\documentclass[aps,prl,twocolumn,superscriptaddress,floatfix,preprintnumbers, nofootinbib,hyperref]{revtex4-2} 
\usepackage{graphicx,amsmath,amssymb,amsfonts,bm,color}
\usepackage{here}
\usepackage{physics}
\usepackage{cases}
\usepackage{hyperref}
\usepackage{natbib}
\begin{document}

\title{Determining the Spin Density Matrix via Its Rank and Probing the Quantum Entanglement and Bell Non-Locality at the Lepton Colliders}

\author{Liwei Liu}
\affiliation{School of Physics, Henan Normal University, Xinxiang 453007, P.R.China}
\author{Xiqing Hao}
\affiliation{School of Physics, Henan Normal University, Xinxiang 453007, P.R.China}
\author{Dianwei Wang}
\affiliation{School of Physics, Henan Normal University, Xinxiang 453007, P.R.China}
\author{Lina Wu}
\affiliation{School of Sciences, Xi’an Technological University, Xi’an 710021, P. R. China}
\author{Tianjun Li}
\email[]{Corresponding author: tli@itp.ac.cn}
\affiliation{School of Physics, Henan Normal University, Xinxiang 453007, P.R.China}

\date{\today}

\begin{abstract}

Considering two-fermion $F_a F_b$ productions and decays via one scalar or photon exchange at the $e^+e^-$ collider, we show that the rank $r_{\rho}$ of spin density matrix $\rho$ is equal to the number of degrees of freedom of the mediator. For one generic scalar exchange, the spin density matrix is rank one for a pure state. With rank-one condition, we can determine the spin analyzing powers for $F_a$ and $F_b$ and their product if the CP symmetry is violated and conserved, respectively, and probe the CP violation. These results can be applied to the $\eta_c \to \Lambda {\bar \Lambda}$ at the BESIII experiment and the Higgs $\to \tau \tau $ at the LHC. For one photon exchange, the spin density matrix is rank two for a mixed state. Considering the  $\Lambda \bar \Lambda $ productions and decays at the BESIII experiment as an example, we show that the spin analyzing powers for $F_a$ and $ F_b$ can be determined by the rank-two conditions in details. Therefore, we can reconstruct the spin density matrices, probe the quantum entanglement and Bell non-locality, and evade the no-go theorem. Furthermore, we conjecture that the $N\times N$ spin density matrix with $r_{\rho} \le N-2 $ can be reconstructed at the lepton colliders in general.

\end{abstract}
	
\maketitle


{\textbf{Introduction.}---}Quantum Mechanics (QM) and special relativity are the cornerstones for modern physics. In QM, there are two characteristic features:
quantum entanglement and Bell non-locality. Quantum entanglement describes a correlation between the sub-systems of a quantum composite system~\cite{Einstein:1935rr,Schrodinger:2008pyl}, which may still hold even if the sub-systems are separated spatially. To distinguish between the QM and Local Hidden Variable Theories (LHVTs), Bell proposed the Bell inequality, which can be satisfied by the LHVTs but violated by QM~\cite{Bell:1964kc, Clauser:1969ny, Horodecki:1995nsk}. The violation of Bell inequality is called Bell non-locality.
During the past decades, quantum entanglement and Bell non-locality not only have been confirmed in the low energy systems, for instance, photons, atoms, and solid state qubits, etc~\cite{Freedman:1972zza, Clauser:1978ng, Aspect:1981nv, Aspect:1981zz, Aspect:1982fx,Weihs:1998gy,Hagley:1997bob, Genovese:2005nw, Yin:2017ips, BIGBellTest:2018ebd}, but also have been applied in the practical fields, for example,  quantum computing~\cite{Jozsa:2002rcj} and quantum communication~\cite{Curty:2003ybi}, etc.

The foundation of the Standard Model (SM) in particle physics is Quantum Field Theory (QFT). Because QFT is based on QM and special relativity, we can probe the
quantum entanglement and Bell non-locality in high-energy collider experiments that are different from the previous low energy experiments. The relativistic scattering and decay processes  at the colliders can generate a lot of entangled states, and thus we can study many systems, such as the top quark pairs, bottom quark pairs, 
 light quark pairs, $\tau$ lepton pairs, $W^{\pm}$ and $Z$ gauge boson pairs,  $\Lambda$ baryons, as well as the particles with different spins, etc~\cite{Tornqvist:1980af, Privitera:1991nz, Abel:1992kz, Dreiner:1992gt, Benatti:1997fr, Benatti:1999jt, Benatti:1999du, Bertlmann:2001ea,
Go:2003tx, Banerjee:2014vga, Acin:2000cs, Li:2008dk, Baranov:2008zzb, Chen:2013epa, Qian:2020ini, Banerjee:2015mha, Yongram:2013soa, Cervera-Lierta:2017tdt, Afik:2020onf, ATLAS:2023fsd, CMS:2024pts, Fabbrichesi:2021npl, Aoude:2022imd,Afik:2022dgh,Fabbrichesi:2022ovb, Fabbrichesi:2023idl,
Ehataht:2023zzt,Barr:2021zcp,Barr:2022wyq,Aguilar-Saavedra:2022wam,Fabbrichesi:2023cev,Severi:2021cnj,Larkoski:2022lmv,Aguilar-Saavedra:2022uye,Afik:2022kwm,Gong:2021bcp,Aguilar-Saavedra:2023hss,Aguilar-Saavedra:2024fig,White:2024nuc,Han:2024ugl,Fabbrichesi:2025ywl,Ashby-Pickering:2022umy,Aguilar-Saavedra:2022mpg, Altakach:2022ywa,Aoude:2023hxv,Morales:2023gow,Bernal:2023ruk,Bi:2023uop,Dong:2023xiw,Ma:2023yvd,Sakurai:2023nsc,Bernal:2023jba,Han:2023fci,Cheng:2023qmz,Aguilar-Saavedra:2024hwd,Aguilar-Saavedra:2024vpd,Aguilar-Saavedra:2024whi,Duch:2024pwm,Morales:2024jhj,Subba:2024mnl,Maltoni:2024csn,Afik:2025grr,Wu:2024asu,Cheng:2024btk,Gabrielli:2024kbz,Ruzi:2024cbt,Cheng:2024rxi,Wu:2024ovc,Ruzi:2024iqu,Altomonte:2024upf,Fabbrichesi:2024rec,Cheng:2025cuv,Han:2025ewp,Guo:2026yhz,Han:2023fci,Bernal:2024xhm,DelGratta:2025qyp,Goncalves:2025mvl,Ruzi:2025jql,Hong:2025drg,Goncalves:2025xer,Altakach:2022ywa,LoChiatto:2024dmx,Ruzi:2024cbt,Ding:2025mzj,Pei:2025yvr,Pei:2025ito,Cao:2025qua,Cheng:2025zcf,Shi:2016bvo,Shi:2019mlf,Shi:2019kjf, Pei:2025non,
Pei:2026rlh,Pei:2026wfu,Pei:2026khg,Antozzi:2026vdi, Barr:2024djo, Fabbrichesi:2024wcd, Subba:2024aut, Zhang:2025mmm, Lu:2025hwy, vonKuk:2025kbv,
Zhang:2026nwm, Yang:2026uwu, Aguilar-Saavedra:2026rsx, Fang:2026ddi, Arai:2026jtc, Roberts:2026hxr, Nguyen:2026amq,
Grossi:2024jae,DelGratta:2025xjp,Pelliccioli:2026ltl}.

 However, in the current collider experiments,  we cannot measure the spins directly, and choose the measurement settings. And then  we cannot  directly measure the spin correlations and spin polarizations in the spin density matrix at the colliders. The main idea of the previous studies is that we can measure the spin correlations indirectly from the proper angular correlations between the final states from the intermediate particle decays, and measure the spin polarizations of the intermediate particles indirectly from the angular distributions of the final states at the colliders.  Especially, the spin-analyzing powers are the key connections between the spin density matrix and the angular distributions/correlations. In order to be logically consistent, we must conduct the spin-analyzing power measurements without the QM and QFT assumptions, which is a great challenge for the quantum entanglement and Bell non-locality studies at the colliders. Therefore, the no-go theorem has been proposed~\cite{Abel:1992kz, Dreiner:1992gt, Bechtle:2025ugc, Abel:2025skj, Li:2024luk, Low:2025aqq}.

Recently, assuming that the spin is defined via the Lorentz symmetry, or considering the implicit symmetry in the spin density matrix, we showed that the trace  ${\rm Tr} [C]$ of the spin correlation matrix $C$ is an invariant quantity~\cite{Wang:2026nls}. Thus, for the exchanges of one mediator, for example, a scalar or a gauge boson,  we can determine the product of the spin-analyzing powers, and reconstruct the spin correlation matrix. With the CHSH-Horodecki criterion~\cite{Clauser:1969ny, Horodecki:1995nsk}, we can test the Bell non-locality, and evade the no-go theorem~\cite{Wang:2026nls}. Therefore, the interesting question is whether we can probe the quantum entanglement and Bell non-locality, and evade the no-go theorem in another brand new approach.

In this paper, we consider two-fermion $F_a F_b$ productions and decays via one scalar or photon exchange at the $e^+e^-$ collider. If  $e^+$ and $e^-$ are random polarized,
 the rank $r_{\rho}$ of spin density matrix $\rho$ for $F_a$ and $F_b$ is equal to the rank of spin density matrix for the mediator, {\it i.e.}, the number of degrees of freedom of the mediator. We discuss the correlations between the spin density matrices and the angular distributions/correlations, and point out that the spin-analyzing powers $\alpha_{F_a}$ and $\alpha_{F_a}$ respectively for $F_a$ and $F_b$ are the only unknown parameters, which need to be determined by the properties of spin density matrices and the experimental measurements. For one generic scalar exchange, the spin density matrix is rank one for a pure state. With rank-one condition, we can determine the product of the spin analyzing powers $F_a F_b$ if the CP symmetry is conserved, and determine the spin analyzing powers for  $F_a$ and $ F_b$ if the CP symmetry is violated. Thus, we can reconstruct the spin density matrix, as well as probe the quantum entanglement, Bell non-locality, and CP violation. Our studies can be applied to the $\eta_c \to \Lambda {\bar \Lambda}$ at the BESIII experiment and the Higgs $\to \tau \tau $ at the LHC. For one photon exchange, the spin density matrix is rank two for a mixed state. Considering the $\Lambda \bar \Lambda $ productions and decays at the BESIII experiment as an example, we show that the spin analyzing powers for $F_a$ and $ F_b$ can be determined by the rank-two conditions in details. Therefore, we can reconstruct the spin density matrix, probe the quantum entanglement and Bell non-locality, and evade the no-go theorem. Furthermore, we only have two unknown parameters $\alpha_{F_a}$ and $\alpha_{F_a}$, and then need two independent equations to determine them. Therefore, we conjecture that the $N\times N$ spin density matrix with $r_{\rho} \le N-2 $ can be reconstructed at the lepton colliders in general.

{\textbf{Spin Density Matrix and Its Correlations with Angular Distributions and  Correlations.}---}We consider the production of a pair of fermions  $F_a F_b$ at the $e^+e^-$ collider, {\it i.e.}, $e^{+}e^{-}\rightarrow F_a F_b$, whose subsequently decays are 
\begin{equation}
F_{a/b} \rightarrow f_{a/b, 1} + f_{a/b, 2} + ... + f_{a/b, N}~.~\,
\end{equation}	
\begin{figure}[!ht]
	\centering
    \includegraphics[]{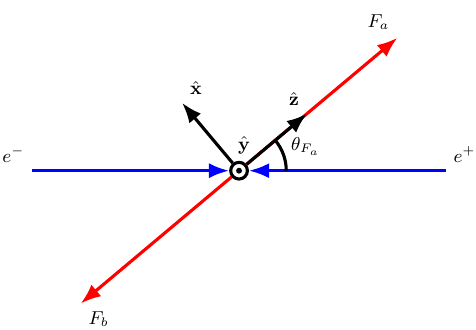}
	\caption{The coordinate system $ (\hat{\vb x}, \hat{\vb y}, \hat{\vb z}) $ for the production process $e^{+}e^{-}\rightarrow F_a F_b$~\cite{Wang:2026nls}.}
	\label{fig:coordinate}
\end{figure}

In the helicity rest frames for $F_a$ and $F_b$, respectively, we define the same coordinate system for them 
\begin{equation}
	\hat{\mathbf{z}}\equiv \hat{\mathbf{p}}_{F_a},\
	\hat{\mathbf{y}}\equiv\frac{\hat{\mathbf{p}}_{e^-}\times\hat{\mathbf{p}}_{F_a}}{\left|\hat{\mathbf{p}}_{e^-}\times\hat{\mathbf{p}}_{F_a}\right|},\ 
	 \hat{\mathbf{x}}\equiv\hat{\mathbf{y}}\times\hat{\mathbf{z}},\label{eq:helicity-rest-frame}
\end{equation}
which is presented in FIG.~\ref{fig:coordinate}~\cite{Wang:2026nls}. Here, $\hat{\mathbf{p}}_{F_a}$ and $\hat{\mathbf{p}}_{e^-}$ are the unit momentum vectors for $F_a$ and $e^-$, respectively.
Because the difference between the rest frames of $F_a$ and $F_b$ is a pure boost along their momentum directions, this coordinate system is indeed convenient.

We assume that both $F_a$ and $F_b$ are sharp resonances due to their small enough decay widths.
From their decay products, we choose two fermions $f_{a, 1}$ and $f_{b, 1}$, respectively. Also,
we assume that we are inclusive over all final-state momenta except for the unit momentum directions ${\vec q}_{a1}$ and ${\vec q}_{b1}$ for $f_{a, 1}$ and $f_{b, 1}$, respectively. Moreover, we formally define  four-vectors  $q^{\mu}_{a1}$ and $q^{\mu}_{b1}$ with $q^0_{a1} \equiv 1 $ and $q^0_{b1} \equiv 1 $, and the expectation values of their products
 \begin{eqnarray}
 	{\mathcal Q}_{\mu \nu}~\equiv~\langle{q^{\mu}_{a1}  q^{\nu}_{b1}\rangle}~,~\,
 \end{eqnarray}
 which can be measured in the collider experiments.

 In Quantum Information Science (QIS), we describe 
the spin polarization state for the quantum composite $F_a F_b$ system via a spin density matrix, and parametrize it via the Fano-Bloch decomposition~\cite{Fano:1983zz}
\begin{align}  
	\rho = \frac{I_{4} + \sum_{i}\left(B_{i}^{+}\sigma^{i} \otimes I_{2} + B_{i}^{-}I_{2} \otimes \sigma^{i}\right) + \sum_{i, j} C_{ij} \sigma^{i} \otimes \sigma^{j}}{4} 
	\label{rho1}
\end{align}
with $i, j = 1, 2, 3$. Here, $I_N$ is the $N\times N$ identity matrix, 
$\sigma_i$ are the Pauli matrices, the coefficients $B_i^{+}$ and $B_i^{-}$ are respectively the net spin polarizations of $F_a$ and $F_b$, and $C_{ij}$ is the spin correlation matrix. If CP is conserved, we have 
$B_i^{+}=B_i^{-}$ and $ C= C^T$.

Defining  $ \sigma^{\mu} \equiv (\sigma^0, \sigma^1, \sigma^2, \sigma^3) $ with $\sigma^0=I_2$,
we can write the above spin density matrix as 
\begin{equation}
	\rho \equiv \frac{1}{4} {\mathcal T}_{\mu\nu} \sigma^\mu \otimes \sigma^\nu~,~
\end{equation}
where ${\mathcal T}_{\mu\nu}$ are defined as
\begin{eqnarray}
{\mathcal T}_{00}~ \equiv ~1~,~
{\mathcal T}_{i0}~\equiv~B_i^+ ~,~ {\mathcal T}_{0i}~\equiv~ B_i^- ~,~ {\mathcal T}_{ij} ~\equiv~ 
C_{ij}~.~\,
\end{eqnarray}

Moreover, in the basis $(|\frac{1}{2}\rangle_{F_a} \otimes |\frac{1}{2}\rangle_{F_b}$, $|\frac{1}{2}\rangle_{F_a} \otimes |-\frac{1}{2}\rangle_{F_b}$, $|-\frac{1}{2}\rangle_{F_a} \otimes |\frac{1}{2}\rangle_{F_b}$, $|-\frac{1}{2}\rangle_{F_a} \otimes |-\frac{1}{2}\rangle_{F_b})$, we can write the spin density matrix as below~\cite{Werner:1989zz, Horodecki:1997vt}
\begin{align}  
	\rho \equiv \sum^n_{i=1} p_i |\Psi_i \rangle \langle \Psi_i|~,~\sum^n_{i=1} p_i =1 ~,~
	 \label{rho2}
\end{align}
where $n \le 16$, and $p_i$ is a positive real number. In addition, $|\Psi_i \rangle \langle \Psi_i|$ is a pure state, and we define its quantum state (polarization state) $|\Psi_i \rangle$ as~\cite{Pei:2025non}
\begin{align}  
	|\Psi_i \rangle \equiv \sum_{k,j=\pm {1\over 2}}\alpha^i_{k,j} |k\rangle_{F_a} \otimes |j\rangle_{F_b}~,~\,
	\label{psii}
\end{align}  
where the normalization condition is 
\begin{align}  
	\sum_{k,j=\pm {1\over 2}}|\alpha^i_{k,j}|^2 = 1~.  
	\label{Normalization}
\end{align}

Next, we briefly review the correlations between the spin density matrix's ${\mathcal T}_{\mu\nu}$ and angular distributions/correlations $\mathcal Q_{\mu \nu}$. In the QFT, such relations are~\cite{Han:2025ewp, Bechtle:2025ugc} 
\begin{align}  
	{\mathcal T}_{\mu\nu} ~=~ \left(\frac{3}{\alpha_{F_a}}\right)^{\eta(\mu)}
	\left(\frac{3}{\alpha_{F_b}}\right)^{\eta(\nu)} {\mathcal Q}_{\mu \nu}
	~,~\,  
	\label{Tqq-1}
\end{align}
where $\eta(0)=0$, and $\eta(1)=\eta(2) =\eta(3)= 1$.
In particular, we have $-1\le \alpha_{F_{a/b}} \le 1$.

In the LHVTs, with the assumptions in Refs.~\cite{Bechtle:2025ugc, Aguilar-Saavedra:2026rsx, Wang:2026nls}, we can obtain the same relations as above in Eq.~(\ref{Tqq-1}). However, the main difference is that
we now have $-3\le \alpha_{F_{a/b}} \le 3$~\cite{Bechtle:2025ugc}. With such larger ranges for the spin-analyzing powers, one can show that the Bell non-locality cannot be probed at the colliders.
Therefore, the great challenge is how to determine the spin-analyzing powers $\alpha_{F_a}$ and $\alpha_{F_b}$ or its product without the QM and QFT assumptions. In the following, we shall show that we can determine the spin-analyzing powers or their product, and reconstruct the spin density
matrix by its rank and the experimental measurements ${\mathcal Q}_{\mu \nu}$.

{\textbf{Generic Scalar Mediator}---}For one generic scalar mediator, because a generic scalar has one degree of freedom, the spin density matrix is rank one for a pure state. The spin polarization state for the quantum composite system $F_{a} F_{b}$ is~\cite{Pei:2025non}
\begin{align}  
	|\Psi_{GS} \rangle = x |\frac{1}{2}\rangle_{F_a} \otimes |-\frac{1}{2}\rangle_{F_b} + y  |-\frac{1}{2}\rangle_{F_a} \otimes |\frac{1}{2}\rangle_{F_b}
	\label{psii-Scalar}
\end{align}  
with the normalization condition $|x|^2+|y|^2=1$.
So the corresponding spin density matrix is 
\begin{align}  
	\label{eq:Cmn}
	\rho_{GS}
	= 	\mqty(
	0 & 0 & 0 &0 \\
	0 & |x|^2 & x y^* & 0 \\
	0 & x^* y & |y|^2 & 0 \\
	0 & 0 & 0 &0 )~.~
\end{align}
In terms of parameters in Eq.~\ref{rho1}, we can prove 
\begin{eqnarray}
&& B_1^{\pm} =B_2^{\pm}=C_{13}=C_{23}=C_{31}=C_{32}=0~,~ \nonumber \\
&& B_3^-=-B_3^+,  C_{21}=-C_{12}, C_{22}=C_{11}, C_{33}=-1~.~	
\end{eqnarray}
From $C_{33}=-1$, we obtain 
\begin{eqnarray}
\alpha_{F_a} \alpha_{F_b} = -9 {\mathcal Q}_{3 3}~.~
\label{psnp-1}
\end{eqnarray}
And then we can determine the spin density matrix as
\begin{eqnarray}
	C_{ij} ~=~ -{\mathcal Q}_{i j} /{\mathcal Q}_{33}~.~\,
\end{eqnarray}
Thus, we can probe the Bell non-locality via the CHSH-Horodecki 
criterion~\cite{Clauser:1969ny, Horodecki:1995nsk}.
For $B_3^-=-B_3^+\not=0$, we get
\begin{eqnarray}
\frac{\alpha_{F_a}}{\alpha_{F_b}} =- \frac{{\mathcal Q}_{30}}{{\mathcal Q}_{03}}~.~
\label{rsnp-1}
\end{eqnarray}
With Eqs.~\ref{psnp-1} and \ref{rsnp-1}, we obtain 
\begin{eqnarray}
\alpha_{F_a} = \pm 3 \sqrt{\frac{{{\mathcal Q}_{3 3} \mathcal Q}_{30}}{{\mathcal Q}_{03}} }~,~
\alpha_{F_b} = \mp 3 \sqrt{\frac{{{\mathcal Q}_{3 3} \mathcal Q}_{03}}{{\mathcal Q}_{30}} }~.~
\end{eqnarray}
From rank-one condition, we get
\begin{eqnarray}
B_3^{+2}+ C_{11}^2+C_{12}^2=1~.~\,
\label{CP-1}
\end{eqnarray}
Because the spin analyzing powers are determined up to a sign difference, we can reconstruct the spin density matrix 
up to an exchange $F_a \leftrightarrow F_b$. Note that the criteria for quantum entanglement and Bell non-locality are the same for such exchange, we do not have any issue to concern.  In addition, we can obtain the condition for quantum entanglement  from the Peres-Horodeki criteria~\cite{Peres:1996dw, Horodecki:1996nc} 
or the discriminant criteria~\cite{Pei:2025non}
\begin{eqnarray}
|B_3^{\pm}| \not= 1~,~ {\rm or}~  {\mathcal Q}_{3 3} \not= {\mathcal Q}_{03}{\mathcal Q}_{30} ~.~\,
\end{eqnarray}
Because $\rho_{GS}$ is a pure state, one can prove that the criteria for  quantum entanglement are the same as the criterion for Bell non-locality via the Gisin theorem~\cite{Gisin:1991vpb, Pei:2025non}.

Furthermore, if CP is conserved, we obtain $B_3^{\pm}=C_{12}=C_{21}=0$. From Eq.~\ref{CP-1}, we obtain 
both $C_{11}$ and $C_{22}$ are equal to +1 and $-1$, which correspond to the exchanges of the CP-even and CP-odd scalars, respectively~\cite{Wang:2026nls}.

We would like to emphasize that our studies for the generic sclar mediator can be applied to the $\eta_c \to \Lambda {\bar \Lambda}$ at the
BESIII experiment and the Higgs $\to \tau \tau $ at the LHC.

{\textbf{Photon Mediator}---}Because photon has two degrees of freedom, the rank of spin density matrix is two. To explain our proposal in details, we present one example: the $\Lambda \bar{\Lambda}$ pair production 
via the $e^+ e^- \rightarrow \gamma^* \rightarrow J/\psi \rightarrow \Lambda \bar{\Lambda}$ process
as well as their decays via $\Lambda\to p+\pi^-$ and $\bar{\Lambda}\to \bar{p}+\pi^+$
at the BESIII experiment.  So we have
$F_{a}\equiv \Lambda$ and $F_{b}\equiv \bar{\Lambda}$, and choose $f_{a,1}\equiv p$ and $f_{b,1}\equiv {\bar p}$.

However, we do not have the BESIII experimental data, and thus, we consider the spin density matrix from the QFT calculations~\cite{Perotti:2018wxm, Batozskaya:2023rek, Wu:2024asu}, whose ${\mathcal T}_{\mu\nu}$ is
\begin{align}  
	&{\mathcal T}_{\mu\nu} = \frac{1}{1+\alpha_\psi C^2_{\Lambda}} \times \nonumber \\
	& \mqty(
	1+\alpha_\psi   C^2_{\Lambda}  &
	0&
	\beta_\psi S_{\Lambda} C_{\Lambda} &
	0\\
	0&
	S^2_{\Lambda}  &
	0&
	\gamma_\psi S_{\Lambda} C_{\Lambda} \\
	\beta_\psi S_{\Lambda} C_{\Lambda} &
	0&
	-\alpha_\psi S^2_{\Lambda} &
	0\\
	0&
	\gamma_\psi S_{\Lambda} C_{\Lambda} &
	0&
	\alpha_\psi+ C^2_{\Lambda})~,~
\end{align}
where $S_{\Lambda} \equiv \sin\theta_\Lambda$, 
$C_{\Lambda} \equiv \cos\theta_\Lambda$,
 $\alpha_{\psi}\in[-1,+1]$ is the decay parameter for the vector charmonium
$\psi$, as well as $ \beta_\psi \equiv \sqrt{1-\alpha_\psi^2} \sin\Delta\Phi$
and $ \gamma_{\psi} \equiv \sqrt{1-\alpha_{\psi}^{2}}\cos(\Delta\Phi)$
with the relative form factor phase $\Delta\Phi\in(-\pi,+\pi]$. 
So the spin density matrix is
\begin{align}  
	&\rho = \frac{1}{4(1+\alpha_\psi C^2_{\Lambda})} \times \nonumber \\
	& \mqty(
	\Delta_+ (1+ C^2_{\Lambda})  & {\mathcal Z}^* S_{\Lambda} C_{\Lambda} &  {\mathcal Z}^* S_{\Lambda} C_{\Lambda}
	 &
	\Delta_+ S^2_{\Lambda} \\
	 {\mathcal Z} S_{\Lambda} C_{\Lambda} &
	\Delta_- S^2_{\Lambda} &
		\Delta_- S^2_{\Lambda} &
	-{\mathcal Z} S_{\Lambda} C_{\Lambda} \\
	 {\mathcal Z} S_{\Lambda} C_{\Lambda} &
	\Delta_- S^2_{\Lambda} &
	\Delta_- S^2_{\Lambda} &
	-{\mathcal Z} S_{\Lambda} C_{\Lambda} \\
	\Delta_+ S^2_{\Lambda} &
	  -{\mathcal Z}^* S_{\Lambda} C_{\Lambda}
	&-{\mathcal Z}^* S_{\Lambda} C_{\Lambda}&
	 \Delta_+ (1+ C^2_{\Lambda}))~,~
	 \label{RhoGamma}
\end{align}
where $\Delta_{\pm} \equiv 1 \pm \alpha_\psi$, and ${\mathcal Z} \equiv  \gamma_\psi+i\beta_\psi$.
And its eigenvalues $\lambda_i$ are
\begin{equation}
	\lambda_{1,2} = \frac{\Delta_{1,2}}{\Delta_1 + \Delta_2},\quad
	\lambda_{3,4} = 0~,~
\end{equation}
where $\Delta_1 \equiv 1 + \alpha_\psi$, and $\Delta_2 \equiv 1 + \alpha_\psi \cos(2\theta_\Lambda)$.
Assuming that the rank-two conditions are well defined, we require $-1 < \alpha_\psi < 1$, which is consistent with the experimental result $\alpha_\psi =  0.4748 \pm 0.0022 \pm 0.0031$~\cite{BESIII:2022qax}.

From the BESIII experiment, we can measure ${\mathcal Q}_{\mu \nu} (\theta_{\Lambda})$. 
To determine the spin analyzing powers, the ideal choice of \(\theta_\Lambda\) would be the value that maximizes the number of zero elements in \(Q_{\mu\nu}(\theta_\Lambda)\). Theoretically, this optimum is achieved at \(\theta_\Lambda = 0\) and \(\pi\), {\it i.e.}, along the beam axis. However, owing to the decay kinematics of \(\Lambda \to p \pi^-\), the flight direction of the final-state proton is highly collinear with the direction of the parent \(\Lambda\) baryon in the laboratory frame. This leads to very low detection efficiency for \(\Lambda\) baryons produced near the beam axis, as their decay products are likewise emitted at forward or backward angles where detector acceptance and tracking efficiency are significantly reduced. Given the angular acceptance of the BESIII detector at the BEPCII, which covers approximately \(21.3^\circ < \theta_\Lambda < 158.7^\circ\), {\it i.e.}, \(|\cos\theta_\Lambda| < 0.93\).
Thus, we consider $\theta_{\Lambda}=\pi/2$. Among
${\mathcal Q}_{i 0}$, ${\mathcal Q}_{0i}$, and ${\mathcal Q}_{i j}$ from experimental measurements, we only have three non-zero parameters ${\mathcal Q}_{i i}$, which satisfy $ {\mathcal Q}_{33}=-{\mathcal Q}_{2 2}
=\alpha_{\psi} {\mathcal Q}_{1 1}$. And then from Eq.~(\ref{Tqq-1}), 
 among $B_i^{\pm}$ and $C_{ij}$, we only have three non-zero parameters $C_{ii}$, which satisfy $C_{33}=-C_{22}=\alpha_{\psi} C_{11}$. 
Thus, we obtain the following spin density matrix for $\theta_{\Lambda}=\pi/2$ from Eq.~(\ref{rho1})
\begin{align}  
	&\rho (\theta_{\Lambda}=\pi/2) = \frac{1}{4} \times \nonumber \\
	& \mqty(
	1+\alpha_\psi C_{11} & 0 & 0 & 	(1+\alpha_\psi) C_{11} \\
	0 & 1-\alpha_\psi C_{11} & (1-\alpha_\psi)C_{11} & 0 \\
	0 & (1-\alpha_\psi) C_{11} &  1-\alpha_\psi C_{11}&  0 \\
(1+\alpha_\psi) C_{11} & 0 & 0 & 	1+\alpha_\psi C_{11} ),~ \nonumber
\end{align}
whose eigenvalues are 
\begin{eqnarray}
	\lambda'_{1,2}=1-C_{11}~,~\lambda'_{3,4}= 1+(1\pm 2\alpha_\psi) C_{11}~.~
\end{eqnarray}
With rank-two conditions, we obtain $C_{11}=1$, and achieve ${\rm Tr} [C]=1$~\cite{Wang:2026nls} 
for $\theta_{\Lambda}=\pi/2$. Thus, we can determine $\alpha_{\psi}$ and
 the product of the spin analyzing powers
\begin{eqnarray}
\alpha_{\psi}={\mathcal Q}_{33}/{\mathcal Q}_{1 1}~,~
	\alpha_{F_a} \alpha_{F_b} = 9 {\mathcal Q}_{11} 
	~{\rm for }~\theta_{\Lambda}=\pi/2 ~.~\,
	\label{SPAs-I}
\end{eqnarray}
Therefore, for arbitrary scattering angle $\theta_{\Lambda}$, we can reconstruct the spin correlation matrix $C_{ij}$ from the experimental measurements ${\mathcal Q}_{i j}$ as follows
\begin{eqnarray}
	C_{ij} ~=~ {\mathcal Q}_{i j} /{\mathcal Q}_{11}(\theta_{\Lambda}=\pi/2)~.~\,
\end{eqnarray}
 With the CHSH-Horodecki criterion~\cite{Clauser:1969ny, Horodecki:1995nsk}, we can probe the Bell non-locality, and evade the no-go theorem. Because it is similar to the discussion in Ref.~\cite{Wang:2026nls}, we shall not repeat it here.

Since we can reconstruct the spin correlation matrix in general, we obtain  
$C_{12} = C_{21}=C_{23}=C_{32}=0$, $C_{13}=C_{31}$, and ${\rm Tr} [C]=1$. Also, from the experimental measurements, we know ${\mathcal Q}_{10}={\mathcal Q}_{01}={\mathcal Q}_{30}={\mathcal Q}_{03} =0$.
So we have $B_1^{\pm} =B_3^{\pm}=0$. To determine the relation between $B_2^{\pm}$, we consider the $3\times 3$ principle minors PMI of spin density matrix in Eq.~(\ref{RhoGamma}) by removing the I-th row and I-th column. Due to the rank-two conditions, the determinant of PMI is zero. Considering PM1 or PM4, we obtain
\begin{eqnarray}
{\rm det(PM1/PM4)} = -(C_{11}+C_{22}) (B_2^+ - B_2^-)^2/64=0~.~\nonumber
\end{eqnarray}
So we obtain $B_2^+ = B_2^-$. And then we have 
\begin{eqnarray}
	\alpha_{F_a}/\alpha_{F_b} = {\mathcal Q}_{20}/{\mathcal Q}_{02}~.~\,
\end{eqnarray}
Considering Eq.~(\ref{SPAs-I}) and
the negative value of  ${\mathcal Q}_{11}(\theta_{\Lambda}=\pi/2)$, we obain the spin analyzing powers for $F_a$ and $F_b$
\begin{eqnarray}
	\alpha_{F_a} &=& \pm 3\sqrt{{\mathcal Q}_{11}(\theta_{\Lambda}=\pi/2) {\mathcal Q}_{20}/ {\mathcal Q}_{02}}~,~ \\ 
	\alpha_{F_b} &=& \mp 3\sqrt{{\mathcal Q}_{11}(\theta_{\Lambda}=\pi/2) {\mathcal Q}_{02}/ {\mathcal Q}_{20}}~.~
\end{eqnarray}
Similarly, we can reconstruct the spin density matrix up to an exchange $F_a \leftrightarrow F_b$.

Next, we study the the quantum entanglement via the the Peres-Horodeki criteria~\cite{Peres:1996dw, Horodecki:1996nc}. Taking partial transpose of $\rho$, we obtain
\begin{align}  
	&\rho^{\rm T_2} = \frac{1}{4(1+\alpha_\psi C^2_{\Lambda})} \times \nonumber \\
	& \mqty(
	\Delta_+ (1+ C^2_{\Lambda})  & {\mathcal Z} S_{\Lambda} C_{\Lambda} &  {\mathcal Z}^* S_{\Lambda} C_{\Lambda}
	&
	\Delta_- S^2_{\Lambda} \\
	{\mathcal Z}^* S_{\Lambda} C_{\Lambda} &
	\Delta_- S^2_{\Lambda} &
	\Delta_+ S^2_{\Lambda} &
	-{\mathcal Z} S_{\Lambda} C_{\Lambda} \\
	{\mathcal Z} S_{\Lambda} C_{\Lambda} &
	\Delta_+ S^2_{\Lambda} &
	\Delta_- S^2_{\Lambda} &
	-{\mathcal Z}^* S_{\Lambda} C_{\Lambda} \\
	\Delta_- S^2_{\Lambda} &
	-{\mathcal Z}^* S_{\Lambda} C_{\Lambda}
	&-{\mathcal Z} S_{\Lambda} C_{\Lambda}&
	\Delta_+ (1+ C^2_{\Lambda}))~,~\nonumber
\end{align}
whose four eigenvalues of $\rho^{T_2}$ are
\begin{equation}
	\lambda^{T_2}_{1,2} = \frac{\Delta_1 \pm \sqrt{\Delta_2^2 - \Delta_3^2}}{2 (\Delta_1 + \Delta_2)},\quad
	\lambda^{T_2}_{3,4} = \frac{\Delta_2 \pm \sqrt{\Delta_1^2 + \Delta_3^2}}{2 (\Delta_1 + \Delta_2)}, \nonumber
\end{equation}
where $\Delta_3 \equiv \beta_\psi \sin(2\theta_\Lambda)$, and we assume 
$\lambda^{T_2}_{1} \le \lambda^{T_2}_{2}$ and $\lambda^{T_2}_{3}\le \lambda^{T_2}_{4}$. We define
\begin{eqnarray}
\Delta_{\rm PH} \equiv  \Delta_1^2 - \Delta_2^2 + \Delta_3^2=
4 \alpha_\psi S^2_{\Lambda}+ (\alpha^2_\psi+ \beta^2_\psi)
 \sin^2(2\theta_\Lambda). \nonumber
\end{eqnarray}
So we have $\lambda^{T_2}_{1} <0$ and $\lambda^{T_2}_{3} <0 $ for $\Delta_{\rm PH}<0$ and $\Delta_{\rm PH}>0$, respectively. Thus, the quantum entanglement condition from  the Peres-Horodeki criteria~\cite{Peres:1996dw, Horodecki:1996nc} is  $\Delta_{\rm PH}\not=0$. Because $\alpha_\psi$ is positive~\cite{BESIII:2022qax}, we obtain the quantum entanglement condition is $\theta_\Lambda \not= 0$ and $\pi$.

{\textbf{Conclusion.}---}We studied quantum tomography for two-fermion $F_a F_b$ productions and decays via one mediator exchange at the $e^+e^-$ collider.
For one generic scalar exchange, the spin density matrix is rank one for a pure state. With rank-one condition, we can determine the spin analyzing powers 
for $F_a$ and $F_b$ and their product for CP violation and conservation, respectively, and probe the CP violation.
Our results can be applied to the $\eta_c \to \Lambda {\bar \Lambda}$ at the BESIII experiment and the Higgs $\to \tau \tau $ at the LHC. For one photon exchange, the spin density matrix is rank two for a mixed state. Considering the  $\Lambda \bar \Lambda $ productions and decays at the BESIII experiment as an example, we showed that the spin analyzing powers for $F_a$ and $ F_b$ can be determined by the rank-two conditions in details. Therefore, we can reconstruct the spin density matrices, probe the quantum entanglement and Bell non-locality, and evade the no-go theorem.

{\textbf{Acknowledgements.}---}L Wu is supported in part by the Natural Science Basic Research Program of Shaanxi, Grant No. 2024JC-YBMS-039.
TL is supported in part by the National Key Research and Development Program of China Grant No. 2020YFC2201504, by the Projects No. 11875062, No. 11947302, No. 12047503, and No. 12275333 supported by the National Natural Science Foundation of China, by the Key Research Program of the Chinese Academy of Sciences, Grant No. XDPB15, by the Scientific Instrument Developing Project of the Chinese Academy of Sciences, Grant No. YJKYYQ20190049, by the International Partnership Program of Chinese Academy of Sciences for Grand Challenges, Grant No. 112311KYSB20210012, and by the Henan Province Outstanding Foreign Scientist Studio Project, No.GZS2025008. This work was supported by the High Performance Computing Platform of Henan Normal University.

\bibliographystyle{apsrev4-2a}
\bibliography{reference}

@article{Gisin:1991vpb,
    author = "Gisin, N.",
    title = "{Bell's inequality holds for all non-product states}",
    doi = "10.1016/0375-9601(91)90805-I",
    journal = "Phys. Lett. A",
    volume = "154",
    number = "5-6",
    pages = "201--202",
    year = "1991"
}

@article{Peres:1996dw,
    author = "Peres, Asher",
    title = "{Separability criterion for density matrices}",
    eprint = "quant-ph/9604005",
    archivePrefix = "arXiv",
    doi = "10.1103/PhysRevLett.77.1413",
    journal = "Phys. Rev. Lett.",
    volume = "77",
    pages = "1413--1415",
    year = "1996"
}

@article{Horodecki:1996nc,
    author = "Horodecki, Michal and Horodecki, Pawel and Horodecki, Ryszard",
    title = "{On the necessary and sufficient conditions for separability of mixed quantum states}",
    eprint = "quant-ph/9605038",
    archivePrefix = "arXiv",
    doi = "10.1016/S0375-9601(96)00706-2",
    journal = "Phys. Lett. A",
    volume = "223",
    pages = "1",
    year = "1996"
}

@article{Wang:2026nls,
    author = "Wang, Dianwei and Hao, Xiqing and Liu, Liwei and Wu, Lina and Li, Tianjun",
    title = "{Determining the Spin-Analyzing Powers via Invariants of the Spin Correlation Matrices and Probing the Bell Non-Locality at the Lepton Colliders}",
    eprint = "2605.13250",
    archivePrefix = "arXiv",
    primaryClass = "hep-ph",
    month = "5",
    year = "2026"
}

@article{Pelliccioli:2026ltl,
    author = "Pelliccioli, Giovanni and Re, Emanuele",
    title = "{SMEFT effects on spin correlations and entanglement at NLO QCD in di-boson production at hadron colliders}",
    eprint = "2601.09540",
    archivePrefix = "arXiv",
    primaryClass = "hep-ph",
    reportNumber = "COMETA-2026-01, LAPTH-003/26",
    month = "1",
    year = "2026"
}

@article{DelGratta:2025xjp,
    author = "Del Gratta, Morgan and Fabbri, Federica and Grossi, Michele and Maltoni, Fabio and Pagani, Davide and Pelliccioli, Giovanni and Vicini, Alessandro",
    title = "{Z-boson quantum tomography at next-to-leading order}",
    eprint = "2509.20456",
    archivePrefix = "arXiv",
    primaryClass = "hep-ph",
    reportNumber = "COMETA-2025-41",
    doi = "10.1007/JHEP02(2026)056",
    journal = "JHEP",
    volume = "02",
    pages = "056",
    year = "2026"
}

@article{Grossi:2024jae,
    author = "Grossi, Michele and Pelliccioli, Giovanni and Vicini, Alessandro",
    title = "{From angular coefficients to quantum observables: a phenomenological appraisal in di-boson systems}",
    eprint = "2409.16731",
    archivePrefix = "arXiv",
    primaryClass = "hep-ph",
    reportNumber = "COMETA-2024-24, MPP-2024-183, TIF-UNIMI-2024-15",
    doi = "10.1007/JHEP12(2024)120",
    journal = "JHEP",
    volume = "12",
    pages = "120",
    year = "2024"
}

@article{Fano:1983zz,
    author = "Fano, U.",
    title = "{Pairs of two-level systems}",
    doi = "10.1103/RevModPhys.55.855",
    journal = "Rev. Mod. Phys.",
    volume = "55",
    pages = "855--874",
    year = "1983"
}

@article{Perotti:2018wxm,
    author = {Perotti, Elisabetta and F{\"a}ldt, G{\"o}ran and Kupsc, Andrzej and Leupold, Stefan and Song, Jiao Jiao},
    title = "{Polarization observables in $e^+e^-$ annihilation to a baryon-antibaryon pair}",
    eprint = "1809.04038",
    archivePrefix = "arXiv",
    primaryClass = "hep-ph",
    doi = "10.1103/PhysRevD.99.056008",
    journal = "Phys. Rev. D",
    volume = "99",
    number = "5",
    pages = "056008",
    year = "2019"
}

@article{Batozskaya:2023rek,
    author = "Batozskaya, Varvara and Kupsc, Andrzej and Salone, Nora and Wiechnik, Jakub",
    title = "{Semileptonic decays of spin-entangled baryon-antibaryon pairs}",
    eprint = "2302.07665",
    archivePrefix = "arXiv",
    primaryClass = "hep-ph",
    doi = "10.1103/PhysRevD.108.016011",
    journal = "Phys. Rev. D",
    volume = "108",
    number = "1",
    pages = "016011",
    year = "2023"
}

@article{Werner:1989zz,
    author = "Werner, Reinhard F.",
    title = "{Quantum states with Einstein-Podolsky-Rosen correlations admitting a hidden-variable model}",
    doi = "10.1103/PhysRevA.40.4277",
    journal = "Phys. Rev. A",
    volume = "40",
    pages = "4277--4281",
    year = "1989"
}

@article{Horodecki:1997vt,
    author = "Horodecki, Pawel",
    title = "{Separability criterion and inseparable mixed states with positive partial transposition}",
    eprint = "quant-ph/9703004",
    archivePrefix = "arXiv",
    doi = "10.1016/S0375-9601(97)00416-7",
    journal = "Phys. Lett. A",
    volume = "232",
    pages = "333",
    year = "1997"
}

@article{Pei:2025non,
    author = "Pei, Junle and Fang, Yaquan and Wu, Lina and Xu, Da and Biyabi, Mustapha and Li, Tianjun",
    title = "{Quantum Entanglement Theory and Its Generic Searches in High Energy Physics}",
    eprint = "2505.09280",
    archivePrefix = "arXiv",
    primaryClass = "hep-ph",
    month = "5",
    year = "2025"
}

@article{Zhang:2026nwm,
    author = "Zhang, Hong-Wei and Cao, Xu and Feng, Tai-Fu",
    title = "{Manipulating Bell nonlocality and entanglement in polarized electron-positron annihilation}",
    eprint = "2602.10389",
    archivePrefix = "arXiv",
    primaryClass = "hep-ph",
    month = "2",
    year = "2026"
}

@article{Yang:2026uwu,
    author = "Yang, Beizhi and Zhang, Yu and Wang, Zeren Simon and Zhou, Xiaorong",
    title = "{Entanglement measures and Bell-type spin-correlation observables in tau-lepton pairs at the Super Tau-Charm Facility}",
    eprint = "2603.05846",
    archivePrefix = "arXiv",
    primaryClass = "hep-ph",
    month = "3",
    year = "2026"
}

@article{Aguilar-Saavedra:2026rsx,
    author = "Aguilar-Saavedra, J. A. and Casas, J. A. and Moreno, J. M.",
    title = "{Understanding Bell locality tests at colliders}",
    eprint = "2603.19389",
    archivePrefix = "arXiv",
    primaryClass = "hep-ph",
    reportNumber = "IFT-UAM/CSIC-26-33",
    month = "3",
    year = "2026"
}

@article{Fang:2026ddi,
    author = "Fang, Yi-Jing and Bhoonah, Amit and Cheng, Kun and Han, Tao and Liu, Yandong and Zhang, Hao",
    title = "{Spin Correlation and Quantum Entanglement of Fermion Pairs in Transversely Polarized $e^-e^+$ Collisions}",
    eprint = "2604.11887",
    archivePrefix = "arXiv",
    primaryClass = "hep-ph",
    reportNumber = "PITT-PACC-2604",
    month = "4",
    year = "2026"
}

@article{Arai:2026jtc,
    author = "Arai, Masato and Mawatari, Kentarou and Okada, Nobuchika",
    title = "{Disentangling new physics with quantum entanglement in $t\bar{t}$ production at future lepton colliders}",
    eprint = "2604.21332",
    archivePrefix = "arXiv",
    primaryClass = "hep-ph",
    month = "4",
    year = "2026"
}

@article{Roberts:2026hxr,
    author = "Roberts, Avalon and Dougan, Patrick and Oh, Alexander and Shaw, Savanna",
    title = "{Odd Physics Off the Diagonal: Constraining CP-violating SMEFT with Quantum Tomography}",
    eprint = "2604.21857",
    archivePrefix = "arXiv",
    primaryClass = "hep-ph",
    month = "4",
    year = "2026"
}

@article{Nguyen:2026amq,
    author = {Nguyen, Hai-Chau and Tetlalmatzi-Xolocotzi, Gilberto and Pardos, Carmen Diez and G{\"u}hne, Otfried and Kleinmann, Matthias},
    title = "{Optimised Inference of Quantum Phenomena in High-Energy Collider Experiments}",
    eprint = "2604.27130",
    archivePrefix = "arXiv",
    primaryClass = "hep-ph",
    reportNumber = "SI-HEP-2026-10, P3H-26-031",
    month = "4",
    year = "2026"
}

@article{BESIII:2022qax,
    author = "Ablikim, M. and others",
    collaboration = "BESIII",
    title = "{Precise Measurements of Decay Parameters and $CP$ Asymmetry with Entangled $\Lambda-\bar{\Lambda}$ Pairs}",
    eprint = "2204.11058",
    archivePrefix = "arXiv",
    primaryClass = "hep-ex",
    doi = "10.1103/PhysRevLett.129.131801",
    journal = "Phys. Rev. Lett.",
    volume = "129",
    number = "13",
    pages = "131801",
    year = "2022"
}

@article{vonKuk:2025kbv,
    author = "von Kuk, Rebecca and Lee, Kyle and Michel, Johannes K. L. and Sun, Zhiquan",
    title = "{Towards a Quantum Information Theory of Hadronization: Dihadron Fragmentation and Neutral Polarization in Heavy Baryons}",
    eprint = "2503.22607",
    archivePrefix = "arXiv",
    primaryClass = "hep-ph",
    reportNumber = "MIT-CTP 5818, Nikhef 2024-020, DESY-25-034",
    month = "3",
    year = "2025"
}

@article{Subba:2024aut,
    author = "Subba, Amir and Singh, Ritesh K. and Godbole, Rohini M.",
    title = "{Looking into the quantum entanglement in $H\to ZZ^\star$ at LHC within SMEFT framework}",
    eprint = "2411.19171",
    archivePrefix = "arXiv",
    primaryClass = "hep-ph",
    month = "11",
    year = "2024"
}

@article{Clauser:1969ny,
    author = "Clauser, John F. and Horne, Michael A. and Shimony, Abner and Holt, Richard A.",
    title = "{Proposed experiment to test local hidden variable theories}",
    doi = "10.1103/PhysRevLett.23.880",
    journal = "Phys. Rev. Lett.",
    volume = "23",
    pages = "880--884",
    year = "1969"
}

@article{Horodecki:1995nsk,
    author = "Horodecki, R. and Horodecki, P. and Horodecki, M.",
    title = "{Violating Bell inequality by mixed spin-1/2 states: necessary and sufficient condition}",
    doi = "10.1016/0375-9601(95)00214-N",
    journal = "Phys. Lett. A",
    volume = "200",
    number = "5",
    pages = "340--344",
    year = "1995"
}

@article{Jozsa:2002rcj,
    author = "Jozsa, Richard and Linden, Noah",
    title = "{On the role of entanglement in quantum-computational speed-up | Proceedings of the Royal Society of London. Series A: Mathematical, Physical and Engineering Sciences}",
    eprint = "quant-ph/0201143",
    archivePrefix = "arXiv",
    doi = "10.1098/rspa.2002.1097",
    journal = "Proc. Roy. Soc. Lond. A",
    volume = "459",
    number = "Issue 2036",
    pages = "2011--2032",
    year = "2003"
}

@article{Curty:2003ybi,
    author = {Curty, Marcos and Lewenstein, Maciej and L{\"u}tkenhaus, Norbert},
    title = "{Entanglement as a Precondition for Secure Quantum Key Distribution}",
    eprint = "quant-ph/0307151",
    archivePrefix = "arXiv",
    doi = "10.1103/PhysRevLett.92.217903",
    journal = "Phys. Rev. Lett.",
    volume = "92",
    number = "21",
    pages = "217903",
    year = "2004"
}

@article{Schrodinger:2008pyl,
    author = {Schr{\"o}dinger, E.},
    title = "{Discussion of Probability Relations between Separated Systems}",
    doi = "10.1017/S0305004100013554",
    journal = "Math. Proc. Cambridge Phil. Soc.",
    volume = "31",
    number = "4",
    pages = "555--563",
    year = "1935"
}

@article{Ehataht:2023zzt,
    author = {Ehat{\"a}ht, Karl and Fabbrichesi, Marco and Marzola, Luca and Veelken, Christian},
    title = "{Probing entanglement and testing Bell inequality violation with $e^+e^-$ {\textrightarrow} ${\ensuremath{\tau}}^+{\ensuremath{\tau}}^-$ at Belle II}",
    eprint = "2311.17555",
    archivePrefix = "arXiv",
    primaryClass = "hep-ph",
    doi = "10.1103/PhysRevD.109.032005",
    journal = "Phys. Rev. D",
    volume = "109",
    number = "3",
    pages = "032005",
    year = "2024"
}

@article{Maltoni:2024csn,
    author = "Maltoni, Fabio and Severi, Claudio and Tentori, Simone and Vryonidou, Eleni",
    title = "{Quantum tops at circular lepton colliders}",
    eprint = "2404.08049",
    archivePrefix = "arXiv",
    primaryClass = "hep-ph",
    doi = "10.1007/JHEP09(2024)001",
    journal = "JHEP",
    volume = "09",
    pages = "001",
    year = "2024"
}

@article{Han:2025ewp,
    author = "Han, Tao and Low, Matthew and Su, Youle",
    title = "{Entanglement and Bell nonlocality in $\tau^+\tau^-$ at the BEPC}",
    eprint = "2501.04801",
    archivePrefix = "arXiv",
    primaryClass = "hep-ph",
    reportNumber = "PITT-PACC-2412",
    doi = "10.1007/JHEP10(2025)217",
    journal = "JHEP",
    volume = "10",
    pages = "217",
    year = "2025"
}

@article{Cheng:2024rxi,
    author = "Cheng, Kun and Han, Tao and Low, Matthew",
    title = "{Quantum tomography at colliders: With or without decays}",
    eprint = "2410.08303",
    archivePrefix = "arXiv",
    primaryClass = "hep-ph",
    reportNumber = "PITT-PACC-2408",
    doi = "10.1016/j.physletb.2025.139675",
    journal = "Phys. Lett. B",
    volume = "868",
    pages = "139675",
    year = "2025"
}

@article{Cheng:2024btk,
    author = "Cheng, Kun and Han, Tao and Low, Matthew",
    title = "{Optimizing entanglement and Bell inequality violation in top antitop events}",
    eprint = "2407.01672",
    archivePrefix = "arXiv",
    primaryClass = "hep-ph",
    reportNumber = "PITT-PACC-2401",
    doi = "10.1103/PhysRevD.111.033004",
    journal = "Phys. Rev. D",
    volume = "111",
    number = "3",
    pages = "033004",
    year = "2025"
}

@article{Privitera:1991nz,
    author = "Privitera, Paolo",
    title = "{Decay correlations in $e^+ e^- \to\tau^+ \tau^-$ as a test of quantum mechanics}",
    reportNumber = "IEKP-KA-91-08, CERN-PPE-91-166",
    doi = "10.1016/0370-2693(92)90872-2",
    journal = "Phys. Lett. B",
    volume = "275",
    pages = "172--180",
    year = "1992"
}

@article{Fabbrichesi:2022ovb,
    author = "Fabbrichesi, Marco and Floreanini, Roberto and Gabrielli, Emidio",
    title = "{Constraining new physics in entangled two-qubit systems: top-quark, tau-lepton and photon pairs}",
    eprint = "2208.11723",
    archivePrefix = "arXiv",
    primaryClass = "hep-ph",
    doi = "10.1140/epjc/s10052-023-11307-2",
    journal = "Eur. Phys. J. C",
    volume = "83",
    number = "2",
    pages = "162",
    year = "2023"
}

@article{Altakach:2022ywa,
    author = "Altakach, Mohammad Mahdi and Lamba, Priyanka and Maltoni, Fabio and Mawatari, Kentarou and Sakurai, Kazuki",
    title = "{Quantum information and CP measurement in $H\to \tau^+\tau^-$ at future lepton colliders}",
    eprint = "2211.10513",
    archivePrefix = "arXiv",
    primaryClass = "hep-ph",
    doi = "10.1103/PhysRevD.107.093002",
    journal = "Phys. Rev. D",
    volume = "107",
    number = "9",
    pages = "093002",
    year = "2023"
}

@article{Ma:2023yvd,
    author = "Ma, Kai and Li, Tong",
    title = "{Testing Bell inequality through $ h{\rightarrow}\tau\tau $ at CEPC*}",
    eprint = "2309.08103",
    archivePrefix = "arXiv",
    primaryClass = "hep-ph",
    doi = "10.1088/1674-1137/ad62d8",
    journal = "Chin. Phys. C",
    volume = "48",
    number = "10",
    pages = "103105",
    year = "2024"
}

@article{Lu:2025hwy,
    author = "Lu, Peng-Cheng and Si, Zong-Guo and Zhang, Han and Zhang, Xin-Yi",
    title = "{Study $\gamma\gamma\to \tau^+\tau^-$ process including $\tau^+ \tau^-$ spin information in Pb-Pb ultraperipheral collision and at Lepton collider}",
    eprint = "2511.18935",
    archivePrefix = "arXiv",
    primaryClass = "hep-ph",
    month = "11",
    year = "2025"
}

@article{Zhang:2025mmm,
    author = "Zhang, Yulei and Zhou, Bai-Hong and Liu, Qi-Bin and Wu, Tong Arthur and Li, Shu and Han, Tao and Hsu, Shih-Chieh and Low, Matthew",
    title = "{Entanglement and Bell Nonlocality in $\tau^+ \tau^-$ at the LHC using Machine Learning for Neutrino Reconstruction}",
    eprint = "2504.01496",
    archivePrefix = "arXiv",
    primaryClass = "hep-ph",
    month = "4",
    year = "2025"
}

@article{Abel:1992kz,
    author = "Abel, S. A. and Dittmar, M. and Dreiner, Herbert K.",
    title = "{Testing locality at colliders via Bell's inequality?}",
    reportNumber = "OUTP-91-40-P",
    doi = "10.1016/0370-2693(92)90071-B",
    journal = "Phys. Lett. B",
    volume = "280",
    pages = "304--312",
    year = "1992"
}

@inproceedings{Dreiner:1992gt,
    author = "Dreiner, Herbert K.",
    title = "{Bell's inequality and tau physics at LEP}",
    booktitle = "{2nd Workshop on Tau Lepton Physics}",
    eprint = "hep-ph/9211203",
    archivePrefix = "arXiv",
    reportNumber = "PRINT-92-0506 (OXFORD), OUTP-92-28-P",
    month = "10",
    year = "1992"
}

@article{Fabbrichesi:2024wcd,
    author = "Fabbrichesi, M. and Marzola, L.",
    title = "{Quantum tomography with $\tau$ leptons at the FCC-ee: Entanglement, Bell inequality violation, $\sin \theta_W$, and anomalous couplings}",
    eprint = "2405.09201",
    archivePrefix = "arXiv",
    primaryClass = "hep-ph",
    doi = "10.1103/PhysRevD.110.076004",
    journal = "Phys. Rev. D",
    volume = "110",
    number = "7",
    pages = "076004",
    year = "2024"
}

@article{Bell:1964kc,
    author = "Bell, J. S.",
    title = "{On the Einstein-Podolsky-Rosen paradox}",
    reportNumber = "RX-1376",
    doi = "10.1103/PhysicsPhysiqueFizika.1.195",
    journal = "Physics Physique Fizika",
    volume = "1",
    pages = "195--200",
    year = "1964"
}

@article{Aspect:1981zz,
    author = "Aspect, Alain and Grangier, Philippe and Roger, Gerard",
    title = "{Experimental Tests of Realistic Local Theories via Bell's Theorem}",
    doi = "10.1103/PhysRevLett.47.460",
    journal = "Phys. Rev. Lett.",
    volume = "47",
    pages = "460--6443",
    year = "1981"
}

@article{Freedman:1972zza,
    author = "Freedman, Stuart J. and Clauser, John F.",
    title = "{Experimental Test of Local Hidden-Variable Theories}",
    doi = "10.1103/PhysRevLett.28.938",
    journal = "Phys. Rev. Lett.",
    volume = "28",
    pages = "938--941",
    year = "1972"
}

@article{Aspect:1981nv,
    author = "Aspect, Alain and Grangier, Philippe and Roger, Gerard",
    title = "{Experimental realization of Einstein-Podolsky-Rosen-Bohm Gedankenexperiment: A New violation of Bell's inequalities}",
    doi = "10.1103/PhysRevLett.49.91",
    journal = "Phys. Rev. Lett.",
    volume = "49",
    pages = "91--97",
    year = "1982"
}

@article{Aspect:1982fx,
    author = "Aspect, Alain and Dalibard, Jean and Roger, Gerard",
    title = "{Experimental test of Bell's inequalities using time varying analyzers}",
    doi = "10.1103/PhysRevLett.49.1804",
    journal = "Phys. Rev. Lett.",
    volume = "49",
    pages = "1804--1807",
    year = "1982"
}

@article{Hagley:1997bob,
    author = "Hagley, E. and Ma{\^\i}tre, X. and Nogues, G. and Wunderlich, C. and Brune, M. and Raimond, J. M. and Haroche, S.",
    title = "{Generation of Einstein-Podolsky-Rosen Pairs of Atoms}",
    doi = "10.1103/PhysRevLett.79.1",
    journal = "Phys. Rev. Lett.",
    volume = "79",
    number = "1",
    pages = "1--5",
    year = "1997"
}

@article{Yin:2017ips,
    author = "Yin, Juan and others",
    title = "{Satellite-based entanglement distribution over 1200 kilometers}",
    doi = "10.1126/science.aan3211",
    journal = "Science",
    volume = "356",
    number = "6343",
    pages = "aan3211",
    year = "2017"
}

@article{BIGBellTest:2018ebd,
    author = "Abell{\'a}n, C. and others",
    collaboration = "BIG Bell Test",
    title = "{Challenging local realism with human choices}",
    eprint = "1805.04431",
    archivePrefix = "arXiv",
    primaryClass = "quant-ph",
    doi = "10.1038/s41586-018-0085-3",
    journal = "Nature",
    volume = "557",
    number = "7704",
    pages = "212--216",
    year = "2018"
}

@article{Weihs:1998gy,
    author = "Weihs, Gregor and Jennewein, Thomas and Simon, Christoph and Weinfurter, Harald and Zeilinger, Anton",
    title = "{Violation of Bell's inequality under strict Einstein locality conditions}",
    eprint = "quant-ph/9810080",
    archivePrefix = "arXiv",
    doi = "10.1103/PhysRevLett.81.5039",
    journal = "Phys. Rev. Lett.",
    volume = "81",
    pages = "5039--5043",
    year = "1998"
}

@article{Clauser:1978ng,
    author = "Clauser, John F. and Shimony, Abner",
    title = "{Bell's theorem: Experimental tests and implications}",
    doi = "10.1088/0034-4885/41/12/002",
    journal = "Rept. Prog. Phys.",
    volume = "41",
    pages = "1881--1927",
    year = "1978"
}

@article{Genovese:2005nw,
    author = "Genovese, Marco",
    title = "{Research on hidden variable theories: A review of recent progresses}",
    eprint = "quant-ph/0701071",
    archivePrefix = "arXiv",
    doi = "10.1016/j.physrep.2005.03.003",
    journal = "Phys. Rept.",
    volume = "413",
    pages = "319--396",
    year = "2005"
}

@article{Tornqvist:1980af,
    author = "Tornqvist, Nils A.",
    title = "{Suggestion for Einstein-podolsky-rosen Experiments Using Reactions Like $e^+ e^- \to \Lambda \bar{\Lambda} \to \pi^- p \pi^+ \bar{p}$}",
    reportNumber = "HU-TFT-80-22",
    doi = "10.1007/BF00715204",
    journal = "Found. Phys.",
    volume = "11",
    pages = "171--177",
    year = "1981"
}

@article{Benatti:1997fr,
    author = "Benatti, F. and Floreanini, R.",
    title = "{Bell{\textquoteright}s locality and ${\varepsilon}'/{\varepsilon}$}",
    eprint = "hep-ph/9712274",
    archivePrefix = "arXiv",
    doi = "10.1103/PhysRevD.57.R1332",
    journal = "Phys. Rev. D",
    volume = "57",
    number = "3",
    pages = "R1332",
    year = "1998"
}

@article{Benatti:1999jt,
    author = "Benatti, F. and Floreanini, R.",
    title = "{Dissipative contributions to epsilon-prime / epsilon}",
    eprint = "hep-ph/9906272",
    archivePrefix = "arXiv",
    doi = "10.1142/S0217732399001619",
    journal = "Mod. Phys. Lett. A",
    volume = "14",
    pages = "1519--1530",
    year = "1999"
}

@article{Benatti:1999du,
    author = "Benatti, Fabio and Floreanini, Roberto",
    title = "{Direct CP violation as a test of quantum mechanics}",
    eprint = "hep-ph/9912348",
    archivePrefix = "arXiv",
    doi = "10.1007/s100520050692",
    journal = "Eur. Phys. J. C",
    volume = "13",
    pages = "267--273",
    year = "2000"
}

@article{Bertlmann:2001ea,
    author = "Bertlmann, R. A. and Grimus, W. and Hiesmayr, B. C.",
    title = "{Bell inequality and CP violation in the neutral kaon system}",
    eprint = "quant-ph/0107022",
    archivePrefix = "arXiv",
    reportNumber = "UWTHPH-2001-23, UWThPh-2001-23",
    doi = "10.1016/S0375-9601(01)00577-1",
    journal = "Phys. Lett. A",
    volume = "289",
    pages = "21--26",
    year = "2001"
}

@article{Banerjee:2014vga,
    author = "Banerjee, Subhashish and Alok, Ashutosh Kumar and MacKenzie, Richard",
    title = "{Quantum correlations in B and K meson systems}",
    eprint = "1409.1034",
    archivePrefix = "arXiv",
    primaryClass = "hep-ph",
    reportNumber = "UDEM-GPP-TH-14-237, UdeM-GPP-TH-14-237",
    doi = "10.1140/epjp/i2016-16129-0",
    journal = "Eur. Phys. J. Plus",
    volume = "131",
    number = "5",
    pages = "129",
    year = "2016"
}

@article{Acin:2000cs,
    author = "Acin, A. and Latorre, J. I. and Pascual, P.",
    title = "{Three party entanglement from positronium}",
    eprint = "quant-ph/0007080",
    archivePrefix = "arXiv",
    doi = "10.1103/PhysRevA.63.042107",
    journal = "Phys. Rev. A",
    volume = "63",
    pages = "042107",
    year = "2001"
}

@article{Li:2008dk,
    author = "Li, Junli and Qiao, Cong-Feng",
    title = "{New Probabilities of Testing Local Realism in High Energy Physics}",
    eprint = "0812.0869",
    archivePrefix = "arXiv",
    primaryClass = "quant-ph",
    doi = "10.1016/j.physleta.2009.09.057",
    journal = "Phys. Lett. A",
    volume = "373",
    pages = "4311",
    year = "2009"
}

@article{Baranov:2008zzb,
    author = "Baranov, S. P.",
    title = "{Bell's inequality in charmonium decays $\eta_c\to \Lambda\bar{\Lambda}$, $\chi_c\to \Lambda\bar{\Lambda}$ and $J/psi \to \Lambda\bar{\Lambda}$}",
    doi = "10.1088/0954-3899/35/7/075002",
    journal = "J. Phys. G",
    volume = "35",
    pages = "075002",
    year = "2008"
}

@article{Chen:2013epa,
    author = "Chen, Shion and Nakaguchi, Y{\={u}}ki and Komamiya, Sachio",
    title = "{Testing Bell's Inequality using Charmonium Decays}",
    eprint = "1302.6438",
    archivePrefix = "arXiv",
    primaryClass = "hep-ph",
    doi = "10.1093/ptep/ptt032",
    journal = "PTEP",
    volume = "2013",
    number = "6",
    pages = "063A01",
    year = "2013"
}

@article{Qian:2020ini,
    author = "Qian, Chen and Li, Jun-Li and Khan, Abdul Sattar and Qiao, Cong-Feng",
    title = "{Nonlocal correlation of spin in high energy physics}",
    eprint = "2002.04283",
    archivePrefix = "arXiv",
    primaryClass = "quant-ph",
    doi = "10.1103/PhysRevD.101.116004",
    journal = "Phys. Rev. D",
    volume = "101",
    number = "11",
    pages = "116004",
    year = "2020"
}

@article{Banerjee:2015mha,
    author = "Banerjee, Subhashish and Alok, Ashutosh Kumar and Srikanth, R. and Hiesmayr, Beatrix C.",
    title = "{A quantum information theoretic analysis of three flavor neutrino oscillations}",
    eprint = "1508.03480",
    archivePrefix = "arXiv",
    primaryClass = "hep-ph",
    doi = "10.1140/epjc/s10052-015-3717-x",
    journal = "Eur. Phys. J. C",
    volume = "75",
    number = "10",
    pages = "487",
    year = "2015"
}

@article{Yongram:2013soa,
    author = "Yongram, N. and Manoukian, E. B.",
    title = "{Quantum field theory analysis of polarizations correlations, entanglement and Bell's inequality: explicit processes}",
    eprint = "1309.2059",
    archivePrefix = "arXiv",
    primaryClass = "hep-th",
    doi = "10.1002/prop.201200137",
    journal = "Fortsch. Phys.",
    volume = "61",
    pages = "668--684",
    year = "2013"
}

@article{Cervera-Lierta:2017tdt,
    author = "Cervera-Lierta, Alba and Latorre, Jos{\'e} I. and Rojo, Juan and Rottoli, Luca",
    title = "{Maximal Entanglement in High Energy Physics}",
    eprint = "1703.02989",
    archivePrefix = "arXiv",
    primaryClass = "hep-th",
    doi = "10.21468/SciPostPhys.3.5.036",
    journal = "SciPost Phys.",
    volume = "3",
    pages = "036",
    year = "2017"
}

@article{Fabbrichesi:2023idl,
    author = "Fabbrichesi, M. and Floreanini, R. and Gabrielli, E. and Marzola, L.",
    title = "{Bell inequality is violated in B0{\textrightarrow}J/{\ensuremath{\psi}}K*(892)0 decays}",
    eprint = "2305.04982",
    archivePrefix = "arXiv",
    primaryClass = "hep-ph",
    doi = "10.1103/PhysRevD.109.L031104",
    journal = "Phys. Rev. D",
    volume = "109",
    number = "3",
    pages = "L031104",
    year = "2024"
}

@article{ATLAS:2023fsd,
    author = "Aad, Georges and others",
    collaboration = "ATLAS",
    title = "{Observation of quantum entanglement with top quarks at the ATLAS detector}",
    eprint = "2311.07288",
    archivePrefix = "arXiv",
    primaryClass = "hep-ex",
    reportNumber = "CERN-EP-2023-230",
    doi = "10.1038/s41586-024-07824-z",
    journal = "Nature",
    volume = "633",
    number = "8030",
    pages = "542--547",
    year = "2024"
}

@article{CMS:2024pts,
    author = "Hayrapetyan, Aram and others",
    collaboration = "CMS",
    title = "{Observation of quantum entanglement in top quark pair production in proton{\textendash}proton collisions at $\sqrt{s} = 13$ TeV}",
    eprint = "2406.03976",
    archivePrefix = "arXiv",
    primaryClass = "hep-ex",
    reportNumber = "CMS-TOP-23-001, CERN-EP-2024-137",
    doi = "10.1088/1361-6633/ad7e4d",
    journal = "Rept. Prog. Phys.",
    volume = "87",
    number = "11",
    pages = "117801",
    year = "2024"
}

@article{Fabbrichesi:2021npl,
    author = "Fabbrichesi, M. and Floreanini, R. and Panizzo, G.",
    title = "{Testing Bell Inequalities at the LHC with Top-Quark Pairs}",
    eprint = "2102.11883",
    archivePrefix = "arXiv",
    primaryClass = "hep-ph",
    doi = "10.1103/PhysRevLett.127.161801",
    journal = "Phys. Rev. Lett.",
    volume = "127",
    number = "16",
    pages = "161801",
    year = "2021"
}

@article{Aoude:2022imd,
    author = "Aoude, Rafael and Madge, Eric and Maltoni, Fabio and Mantani, Luca",
    title = "{Quantum SMEFT tomography: Top quark pair production at the LHC}",
    eprint = "2203.05619",
    archivePrefix = "arXiv",
    primaryClass = "hep-ph",
    reportNumber = "CP3-22-14",
    doi = "10.1103/PhysRevD.106.055007",
    journal = "Phys. Rev. D",
    volume = "106",
    number = "5",
    pages = "055007",
    year = "2022"
}

@article{Afik:2020onf,
    author = "Afik, Yoav and de Nova, Juan Ram{\'o}n Mu{\~n}oz",
    title = "{Entanglement and quantum tomography with top quarks at the LHC}",
    eprint = "2003.02280",
    archivePrefix = "arXiv",
    primaryClass = "quant-ph",
    doi = "10.1140/epjp/s13360-021-01902-1",
    journal = "Eur. Phys. J. Plus",
    volume = "136",
    number = "9",
    pages = "907",
    year = "2021"
}

@article{Afik:2022dgh,
    author = "Afik, Yoav and de Nova, Juan Ram{\'o}n Mu{\~n}oz",
    title = "{Quantum Discord and Steering in Top Quarks at the LHC}",
    eprint = "2209.03969",
    archivePrefix = "arXiv",
    primaryClass = "quant-ph",
    doi = "10.1103/PhysRevLett.130.221801",
    journal = "Phys. Rev. Lett.",
    volume = "130",
    number = "22",
    pages = "221801",
    year = "2023"
}

@article{Afik:2022kwm,
    author = "Afik, Yoav and de Nova, Juan Ram{\'o}n Mu{\~n}oz",
    title = "{Quantum information with top quarks in QCD}",
    eprint = "2203.05582",
    archivePrefix = "arXiv",
    primaryClass = "quant-ph",
    doi = "10.22331/q-2022-09-29-820",
    journal = "Quantum",
    volume = "6",
    pages = "820",
    year = "2022"
}

@article{Barr:2021zcp,
    author = "Barr, Alan J.",
    title = "{Testing Bell inequalities in Higgs boson decays}",
    eprint = "2106.01377",
    archivePrefix = "arXiv",
    primaryClass = "hep-ph",
    doi = "10.1016/j.physletb.2021.136866",
    journal = "Phys. Lett. B",
    volume = "825",
    pages = "136866",
    year = "2022"
}

@article{Barr:2022wyq,
    author = "Barr, Alan J. and Caban, Pawel and Rembieli{\'n}ski, Jakub",
    title = "{Bell-type inequalities for systems of relativistic vector bosons}",
    eprint = "2204.11063",
    archivePrefix = "arXiv",
    primaryClass = "quant-ph",
    doi = "10.22331/q-2023-07-27-1070",
    journal = "Quantum",
    volume = "7",
    pages = "1070",
    year = "2023"
}
\end{document}